# Imaging faint brown dwarf companions close to bright stars with a small, well-corrected telescope aperture

E. Serabyn, D. Mawet, E. Bloemhof, P. Haguenauer, B. Mennesson, K. Wallace
Jet Propulsion Laboratory, California Institute of Technology, Pasadena, CA 91109

and J. Hickey
Palomar Observatory, California Institute of Technology,
PO Box 200, Palomar Mountain, CA 92060

**Abstract**

We have used our 1.6 m diameter off-axis well-corrected sub-aperture (WCS) on the Palomar Hale telescope in concert with a small inner-working-angle (IWA) phase-mask coronagraph to image the immediate environs of a small number of nearby stars. Test cases included three stars (HD 130948, HD 49197 and HR7672) with known brown dwarf companions at small separations, all of which were detected. We also present the initial detection of a new object close to the nearby young G0V star HD171488. Follow up observations are needed to determine if this object is a bona fide companion, but its flux is consistent with the flux of a young brown dwarf or low mass M star at the same distance as the primary. Interestingly, at small angles our WCS coronagraph demonstrates a limiting detectable contrast comparable to that of extant Lyot coronagraphs on much larger telescopes corrected with current-generation AO systems. This suggests that small apertures corrected to extreme adaptive optics (ExAO) levels can be used to carry out initial surveys for close brown dwarf and stellar companions, leaving followup observations for larger telescopes.

**Subject headings:** instrumentation: adaptive optics, stars: low-mass, brown dwarfs

## I. Introduction

The imaging of faint companions close to bright stars requires very high contrast observations at small angular separations, and steady progress is being made in all of the relevant capabilities: an increasing number of novel coronagraphs to "dim" the bright starlight are emerging (e.g. Guyon 2006), contrast-improvement techniques and algorithms are advancing (e.g. Marois et al. 2008), and scattering by non-ideal telescope optics and atmospheric seeing fluctuations should be much reduced with the next generation of ExAO systems. However, ExAO-level coronagraphic systems on large telescopes (Beuzit et al. 2008; Bouchez et al. 2008; Macintosh et al. 2008; Hodapp et al. 2008) can be quite complex and costly, so early validations of novel high-contrast coronagraphic techniques would be very beneficial.

As pointed out by Serabyn et al. (2007), ExAO performance can readily be provided on any telescope that already has an AO system, simply by reimaging a smaller telescope subaperture onto the AO system's deformable mirror (DM). The resultant subaperture diffraction beam is of course broader than the full aperture beam, but such a "well-corrected subaperture" (WCS) immediately enables the use of coronagraphs in the high-Strehl regime in which their true potential can be seen (Sivaramakrishnan et al. 2001). Moreover, the loss of angular resolution can be mitigated to some extent by using a coronagraph with a small IWA, or by a shift to shorter wavelengths, both possible because of the higher wavefront correction level provided in the subaperture. Moreover, an off-axis subaperture can provide an unvignetted WCS with low diffraction features, and which is an ideal match to the clear round pupils necessary for optimal performance of phase-mask coronagraphs (Rouan et al. 2000; Lloyd et al. 2003; Mawet et al. 2005). Giving up some angular resolution for improved wavefront quality may thus be an acceptable tradeoff in the case of high contrast coronagraphy, and indeed, Haguenauer et al. (2005) suggested that an ExAO-level off-axis WCS could actually provide slightly higher contrast at small off-axis angles than is possible with a full telescope aperture corrected only to the more modest levels available with current AO systems (with typical Strehl ratios ~ 0.5).

An interesting question is then whether, in practice, an existing AO system can provide better coronagraphic small-angle detection capabilities by using it to correct an off-axis telescope subaperture extremely well, or the full telescope aperature more modestly. To explore this question experimentally, we have used our WCS on the Palomar Hale telescope in conjunction with a small-IWA focal-plane phase-mask coronagraph to obtain images of nearby stars in the ExAO regime. Our initial coronagraph was a four quadrant phase-mask (FQPM) coronagraph (Rouan et al. 2000, Riaud et al. 2003), in which alternating phases of 0 and $\pi$ radians are applied in the four quadrants of the focal plane. In use, the stellar point spread function (PSF) is centered on the crosshairs of the FQPM, so that the average focal plane electric field is zero. Our initial observations of binary stars with narrowband rejections in excess of 100:1 have been presented elsewhere (Haguenauer et al. 2005, 2006; Serabyn et al. 2006, 2007). Here we focus on the more challenging case of brown dwarf (BD) companions. Prior use of FQPM coronagraphs on AO-corrected full telescope apertures (e.g. Boccaletti et al 2004, 2008) have achieved only modest contrast at small angles, presumably because of high residual wavefront aberrations and scattering due to the secondary mirror-related blockage. In contrast, with ExAO wavefront correction levels and a clear aperture, our WCS has been able to provide contrast levels sufficient to detect brown dwarfs very close to bright stars. The expected performance of well-corrected apertures will be explored from a theoretical point of view in a subsequent paper; here we concentrate on initial experimental demonstrations and performance.

## II. Observations

The Palomar WCS has been described previously (Haguenauer et al. 2005, 2006; Serabyn et al. 2006, 2007), so only a brief description is provided here. In short, a clear off-axis WCS is provided by inserting a set of relay optics upstream of the Palomar AO system (PALAO). This off-axis relay (OAR) magnifies and shifts an off-axis subaperture pupil onto PALAO's DM, yielding ≈ 10 cm actuator spacing in the selected subaperture. The full-width at half-maximum (FWHM) beamwidth of our Palomar WCS is measured to be 280 mas with the current Lyot stop in place, implying a corresponding (stopped-down) subaperture diameter, D, of ≈ 1.6 m.

Two types of FQPM, both manufactured at JPL, were used for these observations. Both masks were designed for use at the wavelength, λ, of the Br γ filter (2.16 μm). The needed half-wave step was generated by evaporation for the first mask (FQPM 1), and by etching for the second mask used (FQPM 3). All astronomical observations were acquired with these masks installed in Palomar's cryogenic infrared PHARO camera. Using PHARO's $K_s$ filter with these masks resulted in maximum peak-to-peak (constructive peak to highest residual peak) laboratory and stellar rejections of ≈ 40 – 50. As this rejection is less than both the theoretical bandwith-limited rejection of our "monochromatic" masks across the $K_s$ band (≈ 250), and the expected seeing-limited peak-to-peak rejection level (a few hundred to one), the performance must be limited by mask imperfections, presumably lateral discontinuities in the phase step edges across the center of the mask (~ 1-2 μm, compared to a PSF FWHM, ~ Fλ, of 35 μm, where F is the beam focal ratio), leakage through a corresponding central "gap" in the mask, or edge quality.

We present observations of four stars here, summarized in Table 1. Three of the stars (HD 130948, HD 49197 and HR7672), with known brown dwarf companions located within a few arc seconds of the primary, were used as test cases, but the last one (HD171488), from the list of nearby young stars of Wichmann et al. (2003), had no previously known companion. Each observation of a target star was accompanied by an observation of a nearby calibrator star (Table 1), which was used for PSF subtraction. The seeing was typically ~ 1" (as measured by a local MASS-DIMM), and the WCS Strehl ratios were typically ~ 0.9. The plate scale (80 ± 2 mas/pixel) and image orientation (typically ± 2°) were derived from observations of binaries.

Our image calibration procedure began in standard fashion with flat fielding, the subtraction of sky background frames, and replacing bad pixels with surrounding average values.

Due to a relative pointing drift between PHARO and the AO system's optical wavefront sensor, the stellar image typically drifted unacceptably far from the center of the FQPM in ~ 100 sec, resulting in steadily degrading performance until the next peak-up occurred (no active post-AO pointing control loop was available at the time of these observations). Thus, in order to maintain as high a contrast as possible, images with peak to peak stellar rejections < 20 were rejected. The remaining "good" images were then coaligned, based on the centration of the central minimum in the FQPM four-fold-symmetric residuals, and added. The reference star's PSF as seen through the FQPM coronagraph was treated in similar fashion, then scaled by the target/reference flux ratio, and subtracted from the target star PSF. The target and reference star PSFs were typically very similar to each other, as illustrated in Fig. 1a for the case of HR7672 and its calibrator. The PSFs for both stars include many common rather long-lived or "semi-static" speckles, which limit the single-image contrast, but subtract better than random speckles. The improvement upon PSF subtraction can be seen in Fig. 1b, where an improvement of about a factor of 10 is seen in the central region, decreasing to a factor of 3 at larger radii. After PSF subtraction, an azimuthal average of the resultant residual stellar PSF was subtracted from each image, followed by median filtering as in Marois et al. (2008). Alternative image reduction approaches, including cross-diagonal subtraction and centro-symmetric subtraction, were also investigated, but PSF subtraction typically gave somewhat better results for these data, so only PSF-subtracted images are presented here. (We note that true companions were relatively more stable against changes in the data reduction technique than noise speckles.) We now discuss the resultant images for each of the observed stars.

### III. Results

**HD130948:** Fig. 2 shows our resultant WCS PSF-subtracted image of HD130948 next to the earlier image of HD130948 obtained at Gemini North by Potter et al. (2002). The (slightly resolved) BD companion pair is clearly easily detected in our image, with a separation from the primary of 2.45" ± 0.1" at PA ~ 105° +/- 2° degrees, and flux ratio $\Delta K_s = 7.37 \pm 0.1$, all of which are very consistent with the earlier image. In our WCS image, the companion pair is located at a radial offset of ~ 8.75 $\lambda/D$, but the same BD (pair) could have been detected much closer to the primary. As seen in Fig. 2, bright PSF residuals are confined to within a radius of ≈ 2 $\lambda/D$ of the stellar position, while beyond 2 $\lambda/D$, the image largely shows a fairly uniform field of faint speckles. Based on the dimness of the speckles beyond 2 $\lambda/D$, we conclude that this same BD pair could have been detected at radial offsets as small as ~ 2 -3 $\lambda/D$.

**HD49197:** HD49197 has a known BD companion much closer to the primary (at ~ 3.4$\lambda$/D for the WCS). Fig. 3 compares our large-scale image of HD49197 to the earlier discovery image obtained using the full Palomar telescope by Metchev & Hillenbrand (2004). In this comparison, the WCS image may be somewhat more suggestive of a companion than the original 5 m image. Fig. 4 compares the central region of our image to the original discovery images obtained with both the full Palomar and Keck telescopes (Metchev & Hillenbrand 2004), on the same linear scale. In our WCS image, the bright PSF residuals are again confined largely inside ~ 2$\lambda$/D of the center. Excluding HD49197C, which lies beyond the field shown in Fig. 4, the brightest spot in our image beyond 2$\lambda$/D (at radius 0.96" ± 0.1" at PA ~ 77° ± 2°, with $\Delta K_s$ = 8.2 ± 0.2) is fully consistent with the location and relative brightness of HD49197C. The next brightest feature, ≈ half as bright as HD49197C, is located approximately on the opposite side of the center (the "southwestern speckle"). This feature changed markedly with reduction technique, suggesting that it is a speckle, while the HD49197C spot remained in place independent of the reduction technique, suggesting that it is indeed a true source. The southwestern speckle thus represents the limiting detectability at very small radii, and beyond ~ 3$\lambda$/D of the center, there is no ambiguity. Thus, if we were in fact unaware of the actual existence of HD49197C, our current WCS image would clearly have served the same role as the discovery image obtained with the full 5 m telescope – it shows a potential candidate companion, and provides a justification for follow-up observations with a larger telescope. A quantitative comparison of detectivities in the three cases shown in Fig. 4 is presented in the next section.

**HR7672:** The brown dwarf near HR7672 was the most challenging case, because of its small angular offset and high contrast ratio (r= 790 ± 5 mas at PA = 157.5° ± 0.5; $\Delta K$ = 8.62 ± 0.07 mag; Liu et al. 2002). For the WCS, this BD thus lies at 2.8 $\lambda$/D. Fig. 5 shows our WCS image, in which an isolated bright spot (at r = 750 mas ± 80 mas, PA = 155 ± 5, with $\Delta K_s$ = 8.7 ± 0.2 mag) is found to coincide almost exactly with the known location of HR7672B (Liu et al. 2002; Boccaletti et al. 2003). This source is the brightest object in the immediate vicinity of the central 2$\lambda$/D "exclusion" zone, except for the wing of one of the two bright FQPM residuals centered inside the (arbitrarily selected) 2$\lambda$/D circle. Another source was also detected further from the center, at radius ~ 5."5 and PA = 149°, but this source does not appear to share the proper motion of HR7672, so is likely not a companion. (This source is located beyond the published FOV of earlier coronagraphic observations, but seems to be barely discernable in the 2MASS data.)

**HD171488:** In contrast to the other sources, HD171488 had no previously known companions. It is a nearby young (30-50 MYr) G0V star from the list of Wichmann et al. (2003). In our image of this source (Fig. 6) a second source is detected 2.7″ ± 0.1″ NE of HD171488, at PA = 35 ± 2°, with $\Delta K_S$ = 6.4. The only previous close companion survey to include this star that we are aware of is the survey of McCarthy & Zuckerman (2004). As their observations of this star were made in 1995 without AO and with a Lyot coronagraph spot radius of 2.5", this companion would have been just at the inner edge of their region of detectability (and perhaps hidden by a bright "rim" around the mask). Although further observations are necessary to confirm whether the faint source that we detect is a bona fide companion, it is statistically possible, since e.g., the density of sources of comparable brightness (K < 14) within a 1' radius in the 2MASS catalog is only 9 x $10^{-4}$ per square arcsec. Moreover, with $K_S$ = 9.4, its brightness is consistent with either a low-mass M star (Baraffe et al. 2003) or, because the primary is young, a young brown dwarf (~ 30 ± 10 $M_J$ with Teff = 2400 ± 100 K in the Baraffe et al. 2003 cond03 model). Spectroscopy and proper motion measurements will be needed to reach a definitive conclusion in this regard.

## IV. Current Performance, Limitations, and Potential Improvements

With coronagraphic images of HD49197 available from the Palomar WCS as well as the full Palomar and Keck apertures, a comparison of coronagraphic point source detection capabilities is possible. To this end, three radial detectivity curves are plotted in Fig. 7, for which the local noise level, $\sigma$, was taken to be the root mean square (rms) deviation of the pixel counts within annuli of width equal to the FWHM, and $4\sigma$ is taken to be the detection threshhold. As is evident in Fig. 7, even with the rather different aperture diameters, ranging from 1.6 m to ~ 10 m, the detection capabilities at small angular offsets (< 1.2") from the primary star are remarkably similar. Indeed, inside ≈ 1.2", the detection limits of the Palomar WCS and the full Palomar telescope are essentially indistinguishable, even though the full-aperture data involved integrations 4 times longer. Moreover, at very small offsets (< 0.7") the WCS performance is even comparable to that of the much larger Keck telescope! It should also be noted that our phase mask coronagraph could in principle detect close binary companions even within the inner $2\lambda/D$ region, as long as they are bright enough ($\Delta Ks$ < 6 - 7), while the opaque mask coronagraphs have no detection capabilities within their comparable central "blind" zones. On the other hand, larger collecting areas inevitably lead to better performance at large radii.

Of course, this comparison is not exactly one to one, because the images were taken at different epochs, with different seeing conditions, and a different integration time in one case, and available hardware and data reduction techniques evolve. Nevertheless, this initial comparison makes clear that small well-corrected apertures can provide near-neighbor detection capabilities competitive with larger, more poorly corrected apertures. As such, an ExAO-level small aperture (whether a WCS or simply a small telescope) can clearly be used to carry out initial time-intensive faint-companion surveys about as well as much larger telescopes equipped with current-generation AO systems. Large telescopes could then be reserved for confirmation and follow-on and observations, and of course for observations of fainter stars. Intriguingly, this conclusion is already suggested by our initial coronagraphic results, using a system that has much room for improvement.

Our current coronagraphic detection capabilities are limited by several factors, including the intrinsic mask rejection, residual pointing and wavefront errors, the size of the collecting pupil, and the size of the wavefront sensor (WFS) subapertures. Intrinsic mask limitations can arise from manufacturing nonidealities (e.g. ragged quadrant edges or edge shifts between quadrants), or the use of a monochromatic mask for broadband light. Since our best short-timescale peak-to-peak broadband contrast ratios were of order 50 even for a laboratory white-light source, while operation of a monochromatic mask over the entire the $K_s$ band sets a limit of $\sim$ 250, intrinsic mask errors such as quadrant edge shifts or raggedness likely limit the peak-to-peak rejection. On the sky, insufficiently accurate tracking mirrors leading to a significant post-AO non-common path pointing drift ($\sim$ 10 mas/sec) severely limited the time that the star could be kept centered on the FQPM "crosshairs." Finally, while the reduced subaperture size limits both the angular resolution and the SNR, the correspondingly reduced WFS cell size also limits sensitivity, because the photon flux per WFS cell is roughly an order of magnitude lower than for the full aperture case. The minimum observable stellar brightness is increased by the same factor, and slower WFS operation is necessary on fainter stars, a limitation that will however apply to any ExAO system. On the other hand, many nearby stars are bright, so this constraint mainly limits the total number of nearby stars that can be observed with an ExAO system.

All of the limitations listed can be improved upon. In particular, in principle more accurate phase masks can be manufactured, as can more achromatic masks. Different types of masks can also be implemented, such as vortex masks (Foo, Palacios & Swartzlander 2005; Mawet et al. 2005) and/or hybrid phase/opaque masks (Boccaletti et al. 2007) which use a tiny opaque spot to block light from leaking through the exact center of the phase mask. The time the star spends centered on the mask can also be increased significantly by stabilizing the post-AO non-common path pointing drift (this problem has recently been greatly reduced at Palomar). In addition, a number of wavefront sensor improvements are feasible in the ExAO regime, which should reduce off-axis scattered light significantly, including non-common-path speckle suppression (Borde & Traub 2006, Give'on et al. 2007, Wallace et al. 2008), spatial filtering of the wavefront sensor (Poyneer & Macintosh 2004), and predictive AO (e.g. Poyneer et al. 2007). Several of these areas are in fact beginning to be addressed with the WCS, leading to the possibility of significantly improved performance. Indeed, even assuming a chromaticity limited peak-to-peak rejection ratio of ~ 250, somewhat longer integrations, and a rejection improvement of an order of magnitude due to PSF subtraction, contrasts of 9-10 mag should be detectable across the entire field of influence of the DM, to within about $\lambda/2D$ of the center (e.g., ~ 110 mas for the Palomar WCS at H band). This level of performance should directly enable a sensitive small-angle BD companion survey, since, as seen above, brown dwarf companions to solar-like main sequence stars occur at contrasts of ~ 8 mag.

Of course the subaperture size can be increased by moving to a larger telescope, which would provide significant benefits, including better angular resolution and sensitivity. Indeed, many modern 8-10 m class telescopes have fractionally smaller secondary blockages than the Hale telescope, thus allowing fractionally larger subapertures. In e.g., the case of a Keck telescope, circular subapertures ~ 4 m diameter are possible, thus improving angular resolution over the Palomar WCS by a factor of ~ 2.5.

## V. Summary and Perspectives

A 1.6 m WCS on Palomar's Hale telescope has been used in conjunction with a FQPM coronagraph to image the immediate surroundings of four nearby stars, leading to rather good images of three test stars with known brown dwarf companions, and to the detection of one new faint source which has a brightness at a level consistent either with a young brown dwarf or a low mass M star companion. Interestingly, even in this early stage of development, the coronagraphic performance achieved with our WCS already compares favorably to that currently obtainable on much larger telescopes using standard AO correction, implying that a small WCS could be used very effectively to carry out a sensitive survey for BD companions. This is the result of three factors: ExAO-level correction of the incoming wavefronts, use of a coronagraph with a small IWA, and use of a clear circular aperture which is an ideal match to phase-mask coronagraphs. Indeed, with further hardware improvements (specifically, with better masks, pointing stability and wavefront control), it may be possible for a small-IWA coronagraph on a modest-sized ExAO-corrected aperture to provide performance (on bright stars) superior to that of classical coronagraphs installed on much larger telescopes corrected by current generation AO systems. This conjecture will be explored further from a theoretical perspective in a subsequent paper, but preliminary predictions are already available in Haguenauer et al. (2005). In any case, the use of a small-IWA coronagraph clearly allows smaller apertures to be considered for the case of coronagraphic space missions.

For any telescope with an existing AO system, a set of relay optics to generate a WCS should be a fairly straightforward and low-cost endeavor, and so a WCS could supply a fast and cost-effective route to ExAO-level high-contrast coronagraphy at any telescope. Such an ExAO-level WCS then immediately allows development work to begin on many of the necessary novel aspects of next-generation ExAO systems (e.g. speckle reduction techniques, wavefront sensor improvements, etc.), while providing a fore-taste of the performance that may eventually be expected on the larger ExAO-corrected apertures. Compared to the Palomar WCS, a larger WCS on a larger telescope would be a significant step forward, especially considering that shorter wavelengths are enabled by the ensuing Strehl ratio improvement, thus allowing one to maintain angular resolution to some extent. In particular, a WCS immediately enables visible wavelength AO (e.g. Serabyn et al. 2007). Assuming on the other hand H-band operation of a 3.5 - 4 m WCS on an 8 – 10 m telescope, an IWA of ~ $\lambda/D$ would correspond to 80 – 100 mas. Moreover, with 30-40 m class telescopes under consideration, WCS systems may be valuable in helping to define "ultimate" ExAO and coronagraphic approaches and limitations long before very costly systems are developed and deployed. On the other hand, the advantages for high contrast coronagraphy of a very well-corrected clear aperture also suggest the possibility of developing a large, very well-corrected off-axis telescope dedicated to high-contrast observations (e.g. Storey et al. 2006), or of making special efforts to correct extremely well one of the large off-axis segments of a telescope such as the Giant Magellan Telescope (www.gmto.org).

**Acknowledgements**

This work was carried out at the Jet Propulsion Laboratory, California Institute of Technology, under contract with the National Aeronautics and Space Administration, and is based in part on observations obtained at the Hale Telescope, Palomar Observatory, as part of a continuing collaboration between the California Institute of Technology, NASA/JPL, and Cornell University. We thank M. Troy and the JPL AO team, and the staff of the Palomar Observatory for their able and ready assistance.

Table 1

| Star | RA | Dec | Spectral type | V mag | K mag | Observing date | FQPM | Integration time (sec) | Approx. seeing (arcsec) | Airmass |
|---|---|---|---|---|---|---|---|---|---|---|
| **HD130948** | **14 50 15.81** | **+23 54 42.64** | **G2V** | **5.88** | **4.45** | **5/9/06** | **#1** | **67** | **0.9** | **1.04** |
| SAO101299 (cal) | 14 57 11.68 | +16 23 17.26 | G5III | 5.72 | 3.55 | 5/9/06 | #1 | 56 | 1 | 1.06 |
| **HD171488** | **18 34 20.10** | **+18 41 24.23** | **G0V** | **7.4** | **5.85** | **5/9/06** | **#1** | **70** | **1.1** | **1.05** |
| SAO103785 (cal) | 18 30 03.97 | +18 35 17.95 | G6III | 7.59 | 5.18 | 5/9/06 | #1 | 35 | 1.1 | 1.03 |
| **HD49197** | **06 49 21.34** | **+43 45 32.80** | **F5** | **7.33** | **6** | **9/10/06** | **#3** | **380** | **1.2** | **1.23** |
| SAO76668 (cal) | 04 36 13.96 | +21 32 11.56 | G5 | 7.6 | 6 | 9/10/06 | #3 | 360 | 1.2 | 1.06 |
| **HR7672** | **20 04 06.22** | **+17 04 12.62** | **G0V** | **5.8** | **4.39** | **9/10/06** | **#3** | **220** | **1.2** | **1.04** |
| SAO106042 (cal) | 20 22 52.37 | +14 33 03.95 | F8V | 6.17 | 4.9 | 9/10/06 | #3 | 50 | 1.2 | 1.07 |

**Figure Captions**

Figure 1: (a) Comparison of summed PSFs obtained for HR7672 (left) and its calibrator (right). (b) Azimuthally averaged detectivity curves before and after PSF subtraction.

Figure 2: Left: original Gemini discovery image of HD130948 B & C (Potter et al. 2002, ApJL 567, L133). Right: PSF-subtracted image of HD130948 obtained with our 1.6 m WCS/FQPM combination on the Hale telescope, on the same scale. The circle is 600 mas in radius, roughly equal to $2\lambda/D$ for the longest wavelength in the $K_s$ band.

Figure 3: Images of HD49197 A, B & C on the same scale, obtained with, left: the full Palomar telescope (from Metchev & Hillenbrand 2004, ApJ 617, 1330), and, right: our Palomar WCS and a FQPM coronagraph (PSF-subtracted image). The circle in panel b is 600 mas in radius, roughly equal to $2\lambda/D$ for the longest wavelength in the $K_s$ band.

Figure 4: The central region of the HD49197 field on the same scale, for three cases: Left: our Palomar WCS image, as in Fig. 3b. Center: Palomar full aperture image. Right: Keck full aperture image. The latter two are from Metchev & Hillenbrand (2004, ApJ 617, 1330). The circle in panel a is 600 mas in radius, roughly equal to $2\lambda/D$ for the longest wavelength in the $K_s$ band.

Figure 5: PSF-subtracted image of the HR7672 field obtained the Palomar WCS and a FQPM coronagraph. The circle is 600 mas in radius, roughly equal to $2\lambda/D$ for the longest wavelength in the $K_s$ band.

Figure 6: PSF-subtracted image of the HD171488 field obtained with the Palomar WCS and a FQPM coronagraph. In this image, the four waffle-mode spikes have been replaced by local average values. The circle is 600 mas in radius, roughly equal to $2\lambda/D$ for the longest wavelength in the $K_s$ band.

Figure 7: 4-$\sigma$ detectability curves calculated from the three HD49197 images in Fig. 4 for the cases of the Palomar WCS, the full Palomar Hale telescope, and the full Keck telescope.

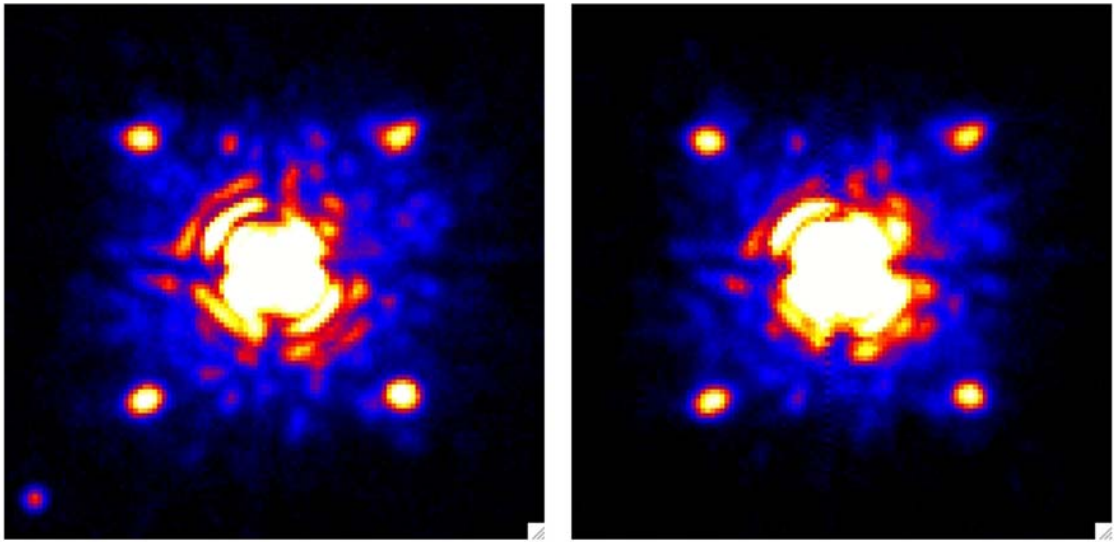

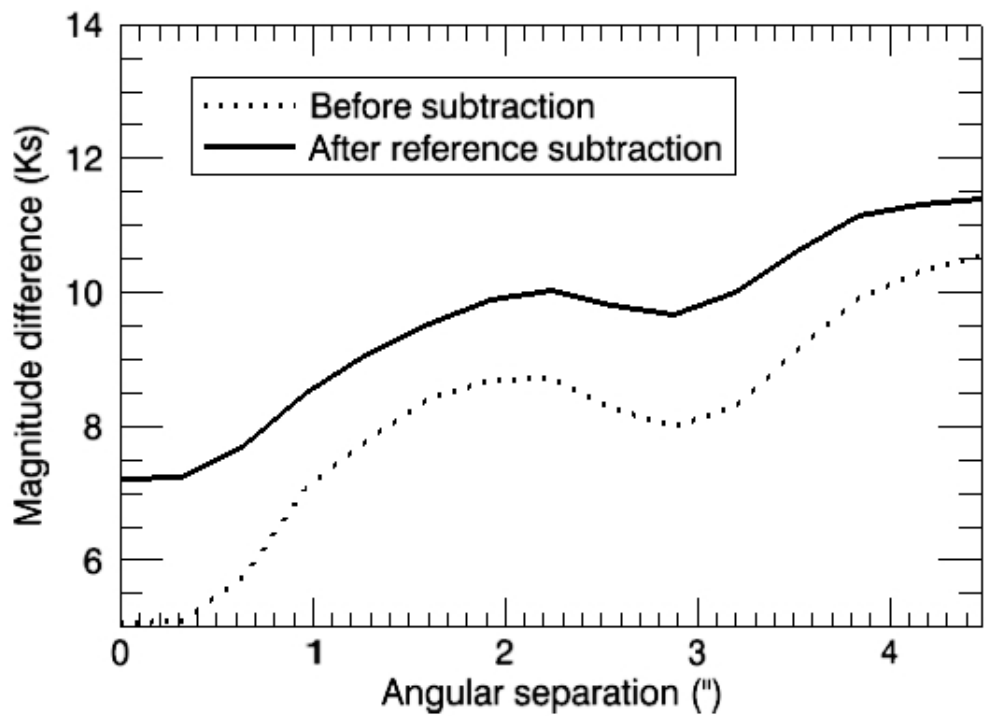

Figure 1: (a) Comparison of summed PSFs obtained for HR7672 (left) and its calibrator (right). (b) Azimuthally averaged detectivity curves before and after PSF subtraction.

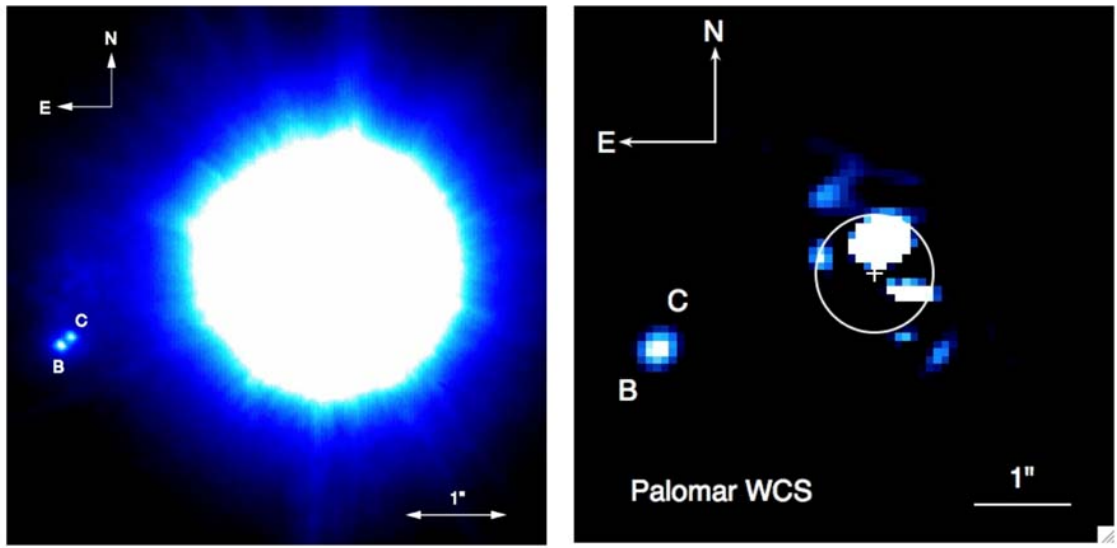

Figure 2: Left: original Gemini discovery image of HD130948 B & C (Potter et al. 2002, ApJL 567, L133). Right: PSF-subtracted image of HD130948 obtained with our 1.6 m WCS/FQPM combination on the Hale telescope, on the same scale. The circle is 600 mas in radius, roughly equal to $2\lambda/D$ for the longest wavelength in the $K_s$ band.

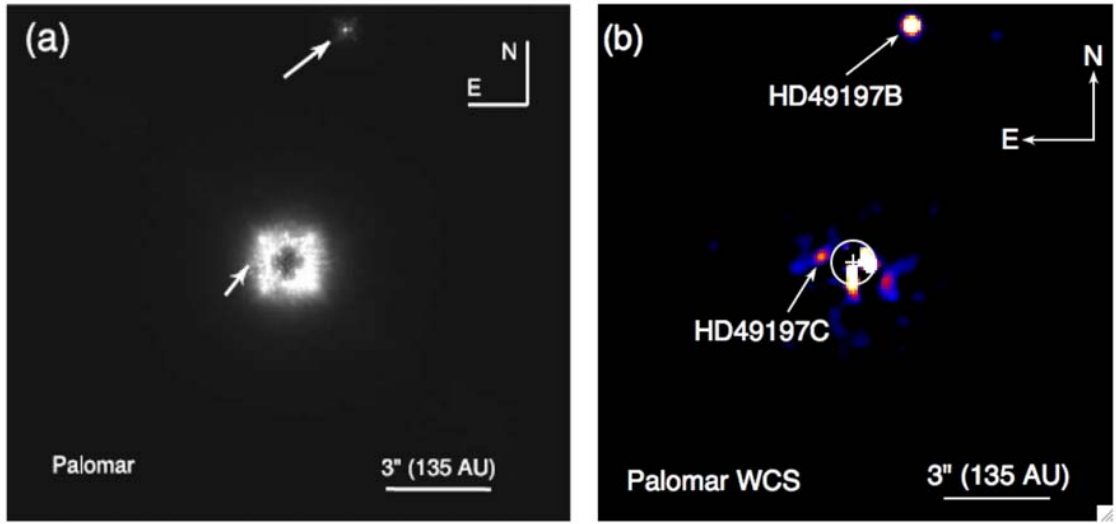

Figure 3: Images of HD49197 A, B & C on the same scale, obtained with, left: the full Palomar telescope (from Metchev & Hillenbrand 2004, ApJ 617, 1330), and, right: our Palomar WCS and a FQPM coronagraph (PSF-subtracted image). The circle in panel b is 600 mas in radius, roughly equal to $2\lambda/D$ for the longest wavelength in the $K_s$ band.

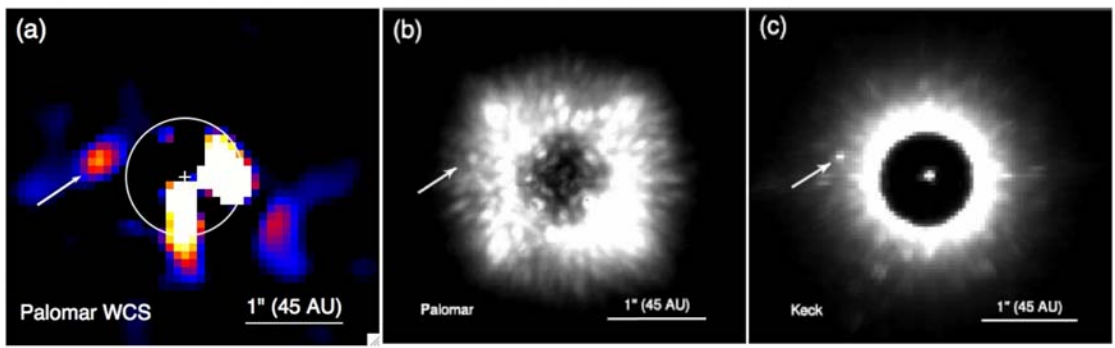

Figure 4: The central region of the HD49197 field on the same scale, for three cases: Left: our Palomar WCS image, as in Fig. 3b. Center: Palomar full aperture image. Right: Keck full aperture image. The latter two are from Metchev & Hillenbrand (2004, ApJ 617, 1330). The circle in panel a is 600 mas in radius, roughly equal to $2\lambda/D$ for the longest wavelength in the $K_s$ band.

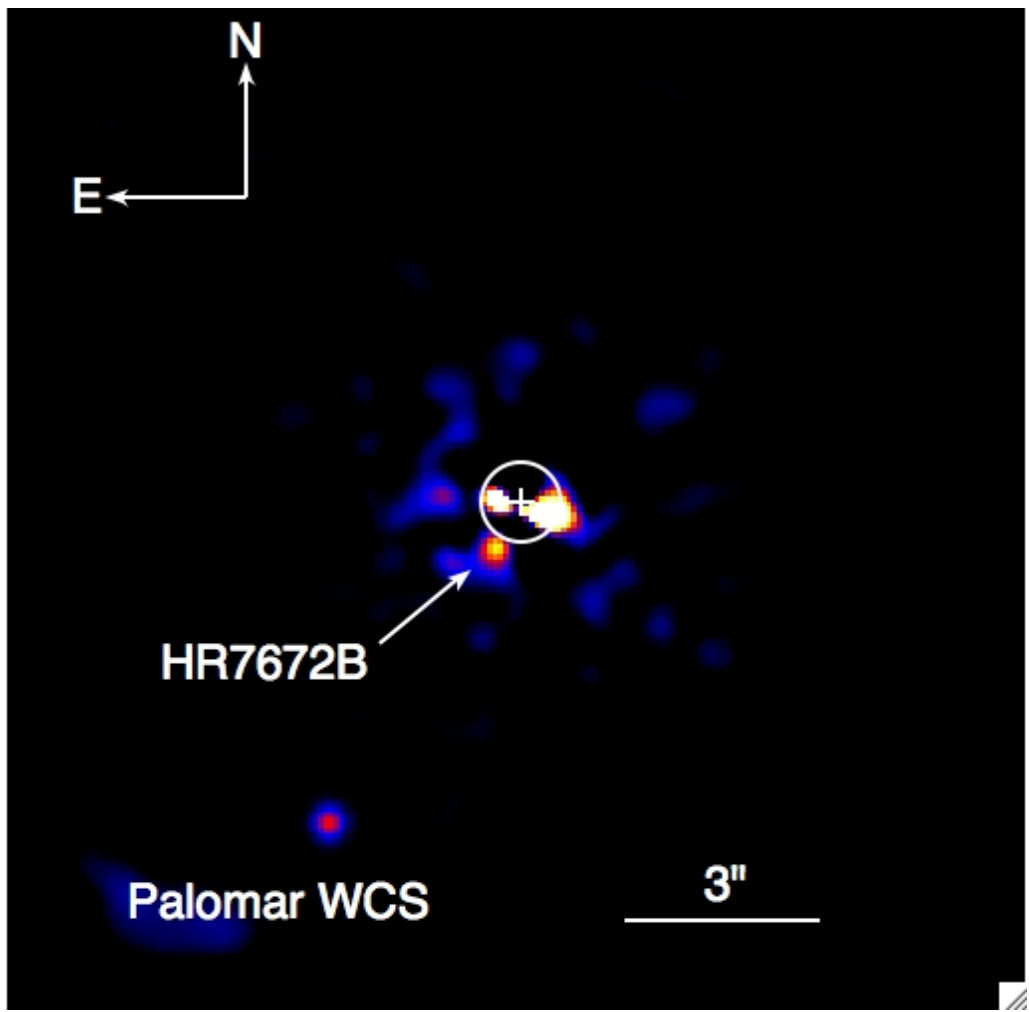

Figure 5: PSF-subtracted image of the HR7672 field obtained the Palomar WCS and a FQPM coronagraph. The circle is 600 mas in radius, roughly equal to 2λ/D for the longest wavelength in the $K_s$ band.

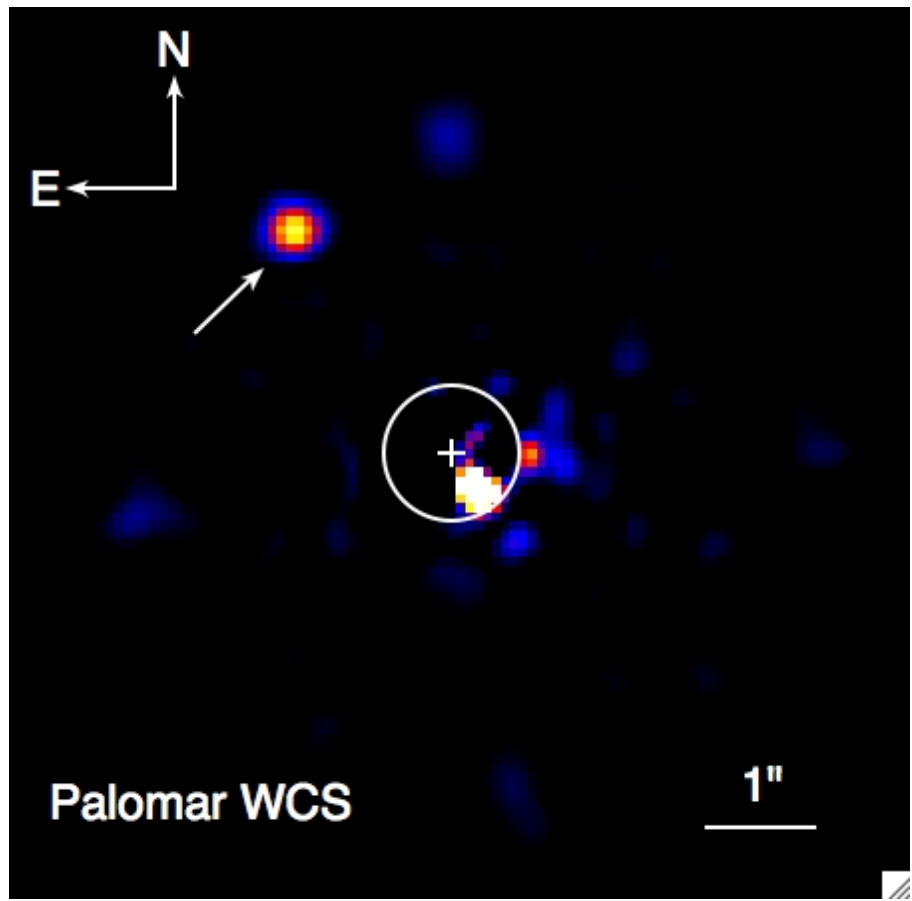

Figure 6: PSF-subtracted image of the HD171488 field obtained with the Palomar WCS and a FQPM coronagraph. In this image, the four waffle-mode spikes have been replaced by local average values. The circle is 600 mas in radius, roughly equal to $2\lambda/D$ for the longest wavelength in the $K_s$ band.

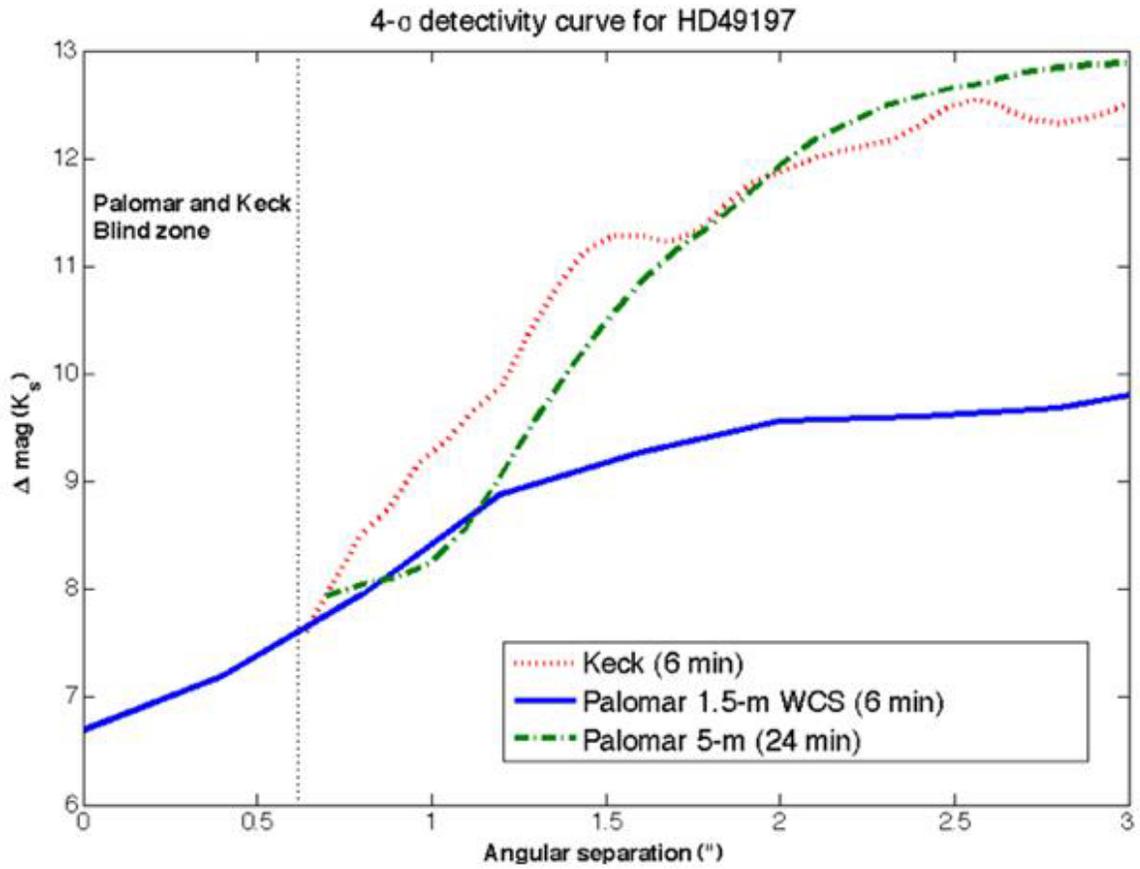

Figure 7: 4-σ detectability curves calculated from the three HD49197 images in Fig. 4 for the cases of the Palomar WCS, the full Palomar Hale telescope, and the full Keck telescope.